\documentstyle{l-aa}
\begin{document}
\thesaurus{02.18.5; 02.19.2; 11.01.2; 11.17.3; 13.25.3}
\title{\bf Inverse Compton X--rays from strong FRII radio--galaxies}
\author{G. Brunetti\inst{1,2} \and G. Setti\inst{1,2} \and A. Comastri\inst{3}}
\institute{Dipartimento di Astronomia, 
via Zamboni 33, I--40126 Bologna, Italy
\and
Istituto di Radioastronomia del CNR, 
via Gobetti 101, I--40126 Bologna, Italy
\and
Osservatorio Astronomico di Bologna, 
via Zamboni 33, I--40126 Bologna, Italy}
\offprints{G. Brunetti, Istituto di Radioastronomia, Via Gobetti 101,
40129 Bologna, Italy}
%
%\date{ Submitted ? 1995; ? 1995 }
\maketitle
%
%\markboth{G. Brunetti et al.}{ }
\begin{abstract}
In the framework of the quasars--radio galaxies unification the 
radio--emitting lobes of FRII radio galaxies are pervaded by an intense
quasar radiation field.
Inverse Compton (IC) scattering between the relativistic electrons 
and the IR--optical photons from a hidden quasar may
provide an important contribution to the X--ray emission of these radio 
galaxies. The soft X--ray emission properties of six strong, high 
redshift FRIIs (3C 277.2, 280, 294, 324, 356, 368) are 
compared with our model expectations, 
taking into account also the  contribution from the IC
scattering of the CMB photons with the radio electrons.
Our estimates are based on a {\it typical} quasar spectrum, derived from the
infrared and optical properties of a 3C quasar sample, and on the assumption of
energy equipartition between relativistic particles and magnetic 
fields  with the same energy density in the electron and proton
components and with a fixed low energy cut--off in the particle distribution
(Appendix A).

We find that the soft X--ray luminosities and
spectra of five out of six sources can be satisfactorily explained by our 
model with the exception of 3C 324 whose X--ray emission is probably dwarfed
by that of the galaxies' cluster of which this source is a member. 

In the case
of 3C 277.2 our model requires a luminosity of the hidden quasar which is
in perfect agreement with that derived from spectropolarimetric studies.
In order to carry out the computations of the IC scattering of the
hidden quasar photons, which are propagating radially outward, we have solved
the anisotropic IC problem. The formal approach and relevant formulae, which
do not appear to be available in the literature, are presented in the
Appendix B. 
One important effect is the prediction that the observed X--ray emission 
associated with the two radio lobes would be asymmetric if the radio
axis is inclined with respect to the plane of the sky, the far-away lobe
being the more luminous. The ratio between the X--ray luminosities of the two
lobes is derived for several values of the inclination of the radio axis.
However, the predicted angular sizes of these distant radio galaxies are small
and difficult to be resolved with present X--ray facilities.
\keywords{Radiation mechanism: non--thermal --- Scattering --- 
Galaxies: active --- quasars: general --- X--rays: general}
\end{abstract}

\section{\bf Introduction}

There is a growing evidence that radio-loud quasars and powerful radio galaxies
with FRII radio structure belong to the same population. According to Barthel
(1989) the quasar phenomenon shows up whenever the line of sight to the source
happen to be within about 45 degrees from the radio axis, while at larger
 viewing angles
the quasar is obscured by a surrounding thick torus of molecular gas co-axial
with the radio axis. Indeed, X-ray observations of the archetypal radio galaxy
Cygnus A have shown that the nuclear source is well described by a heavily
 absorbed ($N_H = 3.7 \times 10^{23}$ cm$^{-2}$) power law spectrum with energy
index $\alpha\simeq 1$ and an
intrinsic isotropic luminosity (2 -- 10 keV) of $\simeq 10^{45}$ erg s$^{-1}$,
 both these figures being typical of the X-ray properties of
steep spectrum radio--loud quasars (Arnaud et al. 1987; 
Ueno et al. 1994). In addition, the recent detection of a broad, but
 relatively weak, Mg II emission
 line typical 
of quasars from this narrow--line radio galaxy (NLRG) 
 can be
 convincingly interpreted as the signature of a hidden quasar whose light is
being scattered into the line of sight by  gas clouds in the host galaxy
(Antonucci et al. 1994). Similarly, the discovery of broad Mg II emission lines
from several NLRGs  indicates the presence 
of a hidden quasar in these sources (di Serego Alighieri et al. 1989; 
Dey \& Spinrad 1996; Cimatti et al. 1993, 96).

A compilation of data obtained with the {\it Einstein} Observatory shows that
the X--ray luminosities of FRII radio galaxies in the 0.3 -- 3.5 keV interval
are comprised in the range $10^{41-45}$ erg s$^{-1}$, with the 
broad--line radio galaxies (BLRG) being
the more luminous ones (Fabbiano et al. 1984), as expected in the unification
model.
ROSAT PSPC pointed observations have now revealed several low redshift BLRGs
whose X--ray emission (unresolved) is consistent with intrinsically absorbed 
spectra (Allen
$\&$ Fabian 1992; Crawford $\&$ Fabian 1995). At one extreme there is the well
known case of 3C 109 (z $\simeq$ 0.3) whose
spectrum is consistent with an absorbed power law  of energy spectral index
$\alpha\sim 0.8$ ($N_H = 5 \times 10^{21}$ cm$^{-2}$)
 and
an unabsorbed luminosity (0.1-2.4 keV) of $ \simeq 5.6 \times 
10^{45}$ erg 
s$^{-1}$.

 While these results generally support the unification scheme
relating the FRII radio galaxies to the radio-loud quasars, nevertheless
detailed studies of the X--ray properties of distant radio galaxies are
 made difficult due to the emission of the hot intracluster gas in which they
might be embedded. Indeed, it has been shown that distant powerful radio 
galaxies 
tend to lie at the centre of moderately rich clusters (Hill $\&$
 Lilly 1991; Yates et al. 1989).

In the case
of Cygnus A, and because of its nearness (z = 0.057), it has been possible to
disentangle part of 
the X--ray structure of the radio galaxy despite the powerful
emission ($\sim 10^{45}$ erg s$^{-1}$) associated with the surrounding hot
cluster gas. ROSAT HRI observations have 
revealed two high brightness emission 
regions, coincident with the radio hot spots, whose intensities
can be explained as Inverse Compton (IC) scattering of the radio photons by the
relativistic electrons in the hot spots themselves
(Harris et al. 1994a).
The evidence for a source centered on the nuclear region of the 
galaxy has been found by subtracting a King model
in the ROSAT HRI frame  (Harris et al. 1994b) and has been confirmed by GINGA
observations (Ueno et al. 1994) revealing 
the presence of an absorbed powerful
AGN type power law in the X--ray spectrum of Cygnus A.
The source can be represented by an unresolved component, but there appears
to be some extended emission present in the measuring circle of radius
$10 ^{\prime\prime}$ (Harris et al. 1994b).
ROSAT HRI observations of the Cygnus A region 
have also revealed significant departures in the X--ray 
brightness distribution from a simple modified King model 
which have been interpreted as the signature of the
hydrodynamic interaction of the jets/lobes with the hot cluster gas 
(Carilli et al. 1994).

The X--ray detections of several others, but
more distant (z $\sim$ 1) powerful NLRGs have been generally interpreted as
due to the emission of a hot intracluster gas either because their spectra do
not show any evidence of intrinsic absorption (3C 356; Crawford  $\&$ Fabian 
1993) or
because by analogy with Cygnus A the absorbed nuclear sources would be too weak
to be detected at large distances (3C 277.2, 368; Crawford  $\&$ Fabian 1995).

In this paper we point out that significant fluxes of X--rays are produced
in the radio lobes of strong radio galaxies by the IC 
process. Recently, Feigelson et al. (1995) have reported the detection of soft
X--rays (ROSAT PSPC) from the radio lobes of the nearby radio galaxy Fornax A
and argued that the surface brightness is consistent with that expected from
the IC scattering of the cosmic microwave background (CMB) 
photons with the relativistic electrons in the lobes
under conditions of energy 
equipartition between particles and
magnetic fields, the relativistic electrons and protons having equal energy
densities. 
These conclusions have been strengthened by the
observations made with ASCA GIS (Kaneda et al. 1995). 
It is well known that the production of X--rays
by this mechanism increases with the redshift since the CMB photon density
is $\propto (1 + z)^{3}$. 
On the other hand, in the framework of the unification
scheme, the radio lobes are pervaded by an intense radiation flux from the
misdirected hidden quasar and  
the ensuing IC losses roughly outweigh those due to the CMB whenever 
$L_{46} \geq  R_{100}^{2}(1 + z)^{4}$,
where $L_{46}$ is the isotropic luminosity of the quasar in units of $10^{46}$
erg s$^{-1}$ and $R_{100}$ is the distance from the quasar in units of 100 Kpc.

In Sect.2 we shall discuss the main assumptions of our model, while the
results of its application to six high redshift radio galaxies, for which 
X--ray data are available, will be presented in Sect.3. The basic IC 
formulae required by our model are derived and discussed in the Appendix B.
In this paper $H_0=75$ km s$^{-1}$ Mpc$^{-1}$ and $q_0=0.0$ are assumed
throughout.

\section{\bf  Model Assumption}
\subsection{\bf Radiation spectrum of the hidden quasars}

Of course one doesn't know what are the radiation properties of the hidden
quasars. We know, however, that quasars' continuum spectra from the far IR 
to the optical can be roughly
approximated by a power law [$F(\nu) \propto \nu^{-\alpha}$] with $\alpha \sim$
 1. From the standpoint of the IC computation the far-- and near--IR
emissions are of particular importance. Based on studies of various
 statistical samples (Sanders et al. 1989; Heckman et al. 1992, 94) 
we have adopted a {\it typical} 
quasar continuum spectrum as a combination of several
power laws: $\alpha$ = 0.2, 0.9, 1.7 and 0.6 respectively in the intervals
100 -- 50$\mu$m, 50 -- 6$\mu$m, 6000 -- 650$nm$ and 650 -- 350$nm$. 
It should be immediately
stressed that a precise knowledge of these quantities is not critical
for the computation of the IC emissivities to be discussed later on.
The {\it typical} spectrum is anchored to the mean rest frame 
luminosity (50--6 $\mu$m)
of the  3CR quasar's sample of Heckman et al. (1994) and to the mean 
optical luminosity (650--350 nm rest frame) for the same 
objects derived from their V
magnitudes (Spinrad et al. 1985) assuming a spectral index  $\alpha$ = 0.6. The 
quasar sample span a redshift range from 0.3 to 2.0 with an average
z $\sim$ 1. We find that the 
integrated isotropic luminosity of the {\it typical} high
redshift radio--loud quasar is $L_{<Q>} = 9.5 \times 10^{46}$ erg s$^{-1}$.
 The corresponding absolute magnitude is
$M_V \sim - 26.1$. We shall refer to the luminosity of this {\it typical}
quasar in our estimates of the IC contribution  to the
 X--ray flux of distant radio galaxies due to the scattering of the photons 
from a hidden quasar. 

In agreement with Barthel's
 unified scheme we adopt the view that
a radio--loud quasar is surrounded by a thick dusty torus
coaxial with the radio structure and with a half opening angle of 45 degrees.
Accordingly we shall assume that only the
relativistic electrons located within the quasar emission cone are those 
involved in the IC scattering of the quasar photons.
However this is strictly the case for the optical photons only. 
Studies of various statistical samples (Heckman et al. 1992, 1994) suggest
that the mean far/near-IR emission of quasars is 4-5 times greater than 
that of FRII radio galaxies, which might be interpreted either as re-radiation
of a sizeable part of the emission from an extended thick dusty torus 
(Pier $\&$ Krolik 1992) 
or as a combination of thermal (isotropic) and non-thermal 
(anisotropic) nuclear radiation (Hes et al. 1995).
Therefore, by assuming the same spatial distribution for both optical and
far/near-IR photons, we shall neglect any possible additional IC contributions 
from electrons located outside the quasar emission cone.

\subsection{\bf Relativistic particles and equipartition fields}

We assume that the radio lobes contain an uniform distribution of
 relativistic electrons and protons with
equal energy density and that there exists an approximate equipartition between
the particle and magnetic field energy densities (minimum energy condition).
The radio lobe spectra can be
generally described by power laws, with typical 
spectral index $\alpha_R$ = 0.8,
 produced by the synchrotron process of ultra--relativistic electrons having
a differential power law energy distribution 
($\propto K_e \gamma^{-\delta}$)
with  $\delta =
 2\alpha_R + 1$.
  The strengths of the equipartition fields, $B_{eq}$, 
are normally computed with
reference to the observed radio flux densities in a given frequency interval, 
say 10 MHz -- 100 GHz source frame (e.g. Pacholczyk 1970).
 Since the synchrotron emission at a
frequency $\nu$ is mainly contributed by the electrons with a Lorentz factor
$\gamma\sim 5\cdot 10^{2} [\nu(MHz)/B_{\perp}(\mu G)]^{1/2}$, 
then the lower bound in this
 frequency interval corresponds to $\gamma$ $= 10^{3}$ -- $10^{4}$ for 
typical
$B_{eq}$= 1 -- 10 $\mu G$ and, as a consequence, a large fraction of the energy
associated with the relativistic particles may reside at lower $\gamma$'s.

On the other hand, electrons with Lorentz factor $\gamma \sim 100-200$ are 
precisely those which are required to produce X--rays via IC scattering of the 
IR--optical photons of the quasar radiation field.
Of course, the modification of the equipartition quantities induced by the
presence of these particles at lower energies may be readily estimated with the
standard equipartition formulae 
by imposing a low frequency cut-off significantly smaller than the 
usually adopted 10 MHz, say 10--100 KHz in 
the source frame.
On the other hand, by choosing the same value of the low frequency cut--off
for sources with different $B_{eq}$ strengths  one would obtain
different lower bound
in the energy distribution of the particles.
In order 
to avoid this formal complication, we have applied equipartition equations 
based on a low energy cut-off ($\gamma_{min}$) 
in the particle distribution (Appendix A).

For a comparison
of the model predicted IC contribution with the observed soft X--ray flux
from a source it
is important to determine the shape of the electron spectrum at low energies
where ionization losses may become relevant. 
In a completely ionized gas the ionization losses 
 equal those due to the combined synchrotron
 and Compton processes at 

\begin{equation}
{\gamma_{\ast}}{\simeq{3.5\cdot10^{2} \sqrt{n_{e}}\over B_{-5}}\,\sqrt{78-
log(n_{e})}} 
\end{equation}

where $n_e$ is the density of thermal electrons and
 $B_{-5} = (B_{eq}^{2} + B_C^{2})^{1/2}$ in units of $10^{-5}$ 
gauss with $B_C$ the
equivalent field for Compton losses. Since the time dependence
of the ionization losses is $\propto \gamma$, by solving the particle energy
continuity equation (Kardashev 1962) one finds that the power law distribution
should flatten by one unit in the exponent, that is $\delta - 1$, for 
$\gamma < \gamma_*$. At large redshifts $B_{-5} > 2.5$ and with
$n_e < 10^{-3}$ cm$^{-3}$ one finds $\gamma_* < 40$. 
From the Faraday 
depolarization studies of high redshift FRIIs with one--sided jets 
Garrington \& Conway (1991) have derived  central values (halo models)
of the $n_eB$ product of about $40 \times 10^{-3}$
cm$^{-3} \mu G$, while
$n_eB < 5 \times 10^{-3}$ cm$^{-3}$ $\mu$G in the radio lobes, under the
assumption that the depolarization is due to thermal gas within the sources.
Then it can be safely assumed that in general $n_e < 5 \times 10^{-4}$
in the radio--lobes. 
As a consequence, we shall adopt
$ \gamma_* =  \gamma_{min}$ = 20
and, accordingly, use Eq.(A3) to compute the equipartition quantities.
If it were $\gamma_* > 20$ the strength of
the equipartition field so derived would be an upper limit, the normalization
of the electron spectrum would be lower (Appendix A),
and the computed X--ray emission would be underestimated
(typically by $\sim 5-15 \%$).
However, if $\gamma_*$ is significantly greater than 150, then the
number of low energy electrons involved in the scattering of the quasar 
IR photons would be too much depleted
and, consequently, our estimates of the IC soft X--ray flux would 
be overestimated.

As a first approximation we assume that the energy density of the relativistic
 particles in the radio lobes remains constant with time.
The spatial distribution of the low energy electrons,
 which mainly contribute
to the soft X--ray flux by the IC scattering of a {\it typical} quasar IR 
radiation field, cannot be directly inferred from radio observations.
From the distribution of the low frequencies brightness profiles 
of powerful radio galaxies 
(Leahy et al. 1989; Carilli et al. 1991), it seems reasonable to assume that 
these particles are distributed approximately with an ellipsoidal geometry 
centered around the galaxy's nuclear region and circumscribing the high
frequency radio lobes, i.e. the site of the ultra relativistic 
particles with  shorter radiative lifetimes.

On the other hand, synchrotron and Compton losses steepen
the power law distribution of the electrons by one unit in the exponent.
Several results (Alexander $\&$ Leahy 1987; Leahy et al. 1989) suggest that,
in general, the corresponding steepening of the radio spectra occurs at 
high frequencies ($\geq 1.5 $GHz), i.e. $ \gamma \geq 10^{4}$ for typical
equipartition fields.
Nevertheless, one can not exclude that for
 the oldest parts of the radio galaxies the high frequency spectral break has 
actually been shifted to very low
 frequencies.
As a consequence for each source we have computed the equipartition quantities 
with reference to the observed flux at low frequencies
(178 MHz), assumed to be emitted by the whole volume describing a
radio source (prolate ellipsoid), in which case the derived synchrotron 
emissivity should be considered as a lower limit, leading to a lower limit on
the normalization of the electron spectrum.

\subsection{\bf Inverse Compton estimates and main model dependences}

First of all, if the radiation field is mainly contributed by the hidden
quasar and the surrounding dusty torus, a very small region compared
with the size of a radio galaxy, then the associated photon momenta are 
anisotropically distributed.
On the other hand
the formula for the IC process have been normally derived 
under the assumption of a locally isotropic
distribution of both electrons and photons momenta
(Blumenthal $\&$ Gould 1970;
Rybicki $\&$ Lightman 1979). Basically, the same results are found in our
model when considering the total 
IC luminosity from a radio galaxy, and the
associated spectral shape, due to the assumed geometry, i.e.  the symmetric 
spatial distribution of the relativistic particles
and the wide half opening angle ($45^{0}$) of the quasar emission cone. 
However, by removing the hypothesis of a locally 
isotropic distribution of the photons'
momenta, the computed IC luminosity 
depends strongly on the angle between the line of sight and the momenta
of the incident photons from the nuclear region.
By solving the anisotropic IC equations it is possible 
to derive the ratio between the X--ray luminosities of the 
two radio lobes. This ratio depends mainly on the
inclination of the radio axis on the plane of the sky, but it also depends
on the assumed geometry (half opening angle of the quasar radiation cone)
and on the spectral energy distribution of the electrons ({\bf Fig.4 \& 5} 
in the Appendix B).
In principle the X--ray luminosity ratio of the two radio lobes is a direct 
observable and it could be combined with the radio depolarization data
to constrain the geometry of the system.

For a given value of the electron spectral slope $\delta$, the IC contribution 
from a hidden quasar to 
the soft X-ray flux of a radio galaxy 
($L^{IC}_{Q}$) is proportional to $K_e$, to the total quasar luminosity
$L_{Q}$ 
and, roughly, to the cubic root of the radio galaxy volume.
A compendium of data (Herbig $\&$ Readhead 1992) shows that the radio 
luminosities
of FRII radio galaxies  span a wide range from $10^{42}$ to 
$10^{45}$ erg s$^{-1}$.
Since optical and (extended) 
radio luminosities of steep spectrum radio quasars appear
to be well correlated (Browne $\&$ Murphy 1987), one can predict that in the 
framework of the unification model the most powerful radio galaxies should
also provide the strongest contribution to $L_Q^{IC}$. 
 
In the case of the hidden quasar model it is straightforward to roughly describe
the expected X-ray brightness profile: while
the quasar photon density decreases with the square of the distance 
($r^{-2}$), the column density of the relativistic electrons along
the line of sight within the quasar emission cone increases with
$r$, so that
the brightness distribution is $\propto r^{-1}$ and consequently the
luminosity of a shell at a  distance $r$ from the nucleus is 
roughly constant.
However, at large distances (several tens Kpc) the quasar emission cone
intercepts the ellipsoidal surface enclosing the  
particle distribution, so that the column density of scattering electrons along
the line of sight starts  decreasing and, as 
a consequence, a systematic decrease of the luminosity profile as a
function of distance is expected (see {\bf Fig.1}).

\begin{figure}
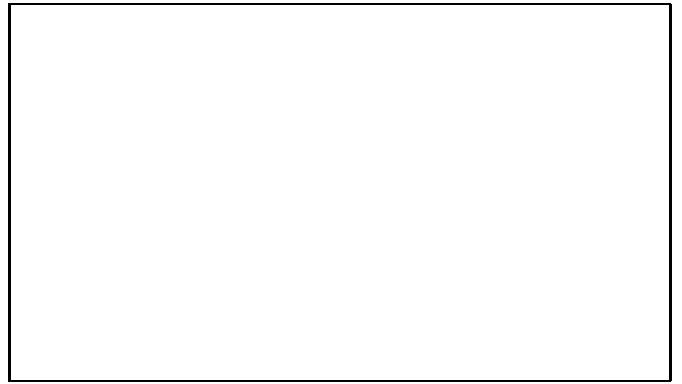

\picplace{5cm}
\caption[]{
The predicted (normalized) IC X--ray luminosity of the radio
galaxy 3C 280, assumed to lie on the sky plane (symmetrical lobes).
The relativistic particles are uniformly distributed in the 
volume bounded by the quasar radiation cone and the ellipsoidal surface
from the 1.4 GHz radio maps. The luminosity profile is obtained by integrating
the surface brightness distribution over circular arcs at a distance
$r$ from the nucleus. The total luminosity profile (solid line) is the sum of 
the contribution from the IC scattering of the hidden quasar photons
(dotted line) and of those of the CMB (dashed line). The contribution of the 
"CMB" to the total luminosity is $\sim$ 10\%.
The radio lobes of the 5 GHz map extend inward to $\sim$ 15 Kpc, while those
of the 1.4 GHz map reach the nuclear region (Ronghui et al. 1992).}
\end{figure}

In addition to the quasar model,
the contribution to the X--ray flux from the IC scattering of the CMB photons
in powerful radio galaxies may become increasingly important with the 
redshift, the IC emissivity being 
$\propto (1+z)^{{\delta+5}\over 2}$ (e.g. Blumenthal \& Gould 1970).
The density of the CMB photons is constant throughout the radio source
volume and the photons' momenta are locally isotropic. Given the much softer 
spectrum of the CMB the relativistic 
electrons responsible for a given X--ray IC emission are on the average more
energetic than those required in the IC scattering of the hidden quasar
radiation field.
The contribution
to the soft X--ray flux from each source has been estimated by applying standard
formulae in the literature (e.g., Harris \& Grindlay 1979) to our source model.

The estimates of the IC fluxes depend, other things being equal, 
on the assumed orientation of
the radio axis.
Let the angle between the line of sight and the radio axis be $\theta_{ax}$
({\bf Fig.4} in the Appendix B).
The volume of a radio galaxies (assumed to be ellipsoidal) is roughly  
$\propto (\sin\theta_{ax})^{-1}$, 
so that from the synchrotron emissivity formula 
and from Eq.(A3)
 it is found that the contribution to the IC flux from the scattering
of the CMB photons is
$\propto (\sin\theta_{ax})^{-{{\delta+1}\over{\delta+5}}}$,
i.e. it increases with decreasing $\theta_{ax}$. 
The dependence on the inclination angle of the IC flux from the
scattering of the quasar radiation field cannot be
 evaluated with simple geometrical considerations. Taking into account the
dilution of the quasar photon field one would find that 
for an elongated ellipsoidal distribution the IC flux roughly 
decreases with $\theta_{ax}$ as $(\sin\theta_{ax})^{{\delta+9}\over
{\delta+5}}$, 
but this is more than compensated by the enhanced IC emission of the far radio
lobe ({\bf Fig.5} in the Appendix B). 
 
While, on the one hand (Sect. 2.2), the adopted source volumes
lead to lower 
limits for the synchrotron emissivity and the normalization of the
electron spectra, on the other hand it might be argued that 
the total number of the 
relativistic particles in the radio galaxies are overestimated, thus leading
to an overestimate of the derived X--ray luminosities.
It should be stressed , however, that this effect might be relevant only in the
case of the IC scattering with the CMB photons, but not
in the case of the IC scattering of the  quasar photons, for which the  
electrons responsible of the scattering are only those located within the
quasar emission cone.
If the particles are uniformly distributed
within an ellipsoidal volume
of minor axis $b$, then  it can be shown with
simple geometrical considerations that, under equipartition conditions,
the luminosity due
to IC scattering of the hidden quasar photons goes as 
$b^{{2\delta-5}\over 7}$, while that of the CMB photons as
$b^{{2(\delta+1)}\over 7}$.
For instance, with $\delta=2.6$ a change of $b$ by a factor 3 (i.e.
a factor 9 in the volume) means a change of a factor $\sim$3 in the
predicted X-ray luminosity from the IC scattering of the CMB photons, but only
$\sim$10 \% in that predicted by the IC scattering of the quasar photons.

It should be noted, however, that the minor axis $b$ in our source models
is not a free parameter.
It is derived by fitting the outer contours of the high-frequency radio lobes
of each source.
Therefore, from the standpoint of the present discussion it might be
more appropriate to consider an alternative situation where particles
and fields are uniformly distributed only in that part of the ellipsoidal 
volume intercepted by the quasar radiation cones.
In this case the reduction in the volume implies an increase in the 
equipartition particle density (eq.A3) and, consequently, an increase in the
predicted X-ray flux from the IC scattering of the quasar photons with respect
to the uniformly filled ellipsoidal volume model.
This, in general,  
more than compensate the decrease in the predicted X-ray luminosity
from the scattering of the CMB photons.

A decrease in the predicted X--ray luminosities is expected if the relativistic
particles are progressively depleted within the quasar radiation cones.
With reference to Fig.1, which illustrates the application of our model
to one of the sources discussed in the next section,
one sees that by emptying the first 5--10 Kpc the expected luminosity would
decrease by 10--20 \%.

\section{\bf Results}

We have applied our model to 
 six high redshift, very powerful 
 FRII  radio galaxies which have been previously 
detected with ROSAT (3C 277.2, 280,
294, 324, 356 and 368). The IC computations have been made 
under the assumption of three different orientations
of the radio axis with respect to the line of sight, that is 90, 60 and
45 degrees. For each configuration we have derived the luminosity of the
hidden quasar which is required to match the observed total soft
X--ray
luminosity of each source and the ratio between the X--ray fluxes of the
two lobes predicted by our model. The source
fluxes at 178 MHz and the radio spectra have been
taken from Laing et al. (1983) and Herbig et al. (1992).
The extension and shape of the ellipsoidal distribution of the
relativistic
particles for each radio galaxy have been inferred from the available
radio-maps.
A detailed discussion for each object is reported in the following,
while the main results are summarized in Table 1.

\subsection{{\bf 3C 277.2} (z=0.776)}

This radio galaxy has been observed  with ROSAT PSPC
for a total exposure time of 13.5 ksec and
 detected only at the 3$\sigma$ level, 
corresponding to about 12 source net counts (Crawford $\&$ Fabian 1995).
Assuming the X--ray spectrum predicted by our IC model,
a power law of energy index 0.95,
the derived 0.2--4.0 keV  
luminosity is 2.6 $\times 10^{43}$ erg s$^{-1}$.
The overall angular extension of the radio structure is $\sim 58$ arcsec and
the radio volume has been modeled (ellipsoid semi--axis $192\times 35$ Kpc) 
by making use of the 1502 MHz map
from 
Pedelty et al. (1989).  
If  3C 277.2 lies on the sky plane $(\theta_{ax} = 90^{o})$, we find
that the 
IC scattering of the CMB photons may contribute up to 22\% of the source
X--ray 
flux, while the required luminosity of a hidden quasar which could
account for 
the remaining part of the observed flux via the IC process would be
0.93 $L_{<Q>}$. 
The predicted total angular size 
of the X--ray source is $\sim$ 20 arcsec for an e--folding drop in the 
luminosity profile. If the radio axis inclination is
$\theta_{ax} = 60^{o}$ the corresponding figures would be 24\% and 0.80
$L_{<Q>}$,
while the
X--ray source would appear asymmetric with one lobe being $\sim$ 3 times 
brighter than the other one. 

The presence of a hidden quasar in 3C 277.2 has already 
been proposed to explain 
the strong radio--optical alignment and the high degree of polarization
of the optical spectrum (di Serego Alighieri 1994 and ref. therein).
From a
recent modeling based on dust scattering it is estimated that the hidden
quasar would have a magnitude
$m_V \simeq 17.2$ (Manzini \& di Serego Alighieri 1996). 
It is interesting to note
that, by
assuming the {\it typical} quasar spectrum, the V magnitude of the hidden
quasar
required by our model  
would be 16.9 and 17.1, for $\theta_{ax} = 90^{o}$ and $60^{o}$ 
respectively, in very
good agreement with the above estimate.

\subsection{{\bf 3C 280} (z=0.996)}

3C 280 is one of the most powerful FRII radio galaxies.
Deep optical imaging revealed that the $[O II]$
emission lines of this radio galaxy  are edge--brightened and
lie along the boundaries of the radio lobes (McCarthy et al. 1995 and
references therein).  
The source is quite faint in the soft X--ray energy band 
giving only about 71 net counts on a deep ROSAT PSPC
exposure of about 48 ksec (Worral et al. 1994).
The spectrum is poorly constrained being consistent
with a power law spectral index $0.5 < \alpha <2.0$  
and an upper limit on the intrinsic absorption
of $4 \times 10^{21}$ cm$^{-2}$ suggesting that the soft X-ray emission
is not
intrinsically absorbed. By adopting a spectral index $\alpha=0.71$,
consistent
with the radio data, we find a luminosity of 6.9 $\times 10^{43}$ erg s
$^{-1}$ in
the 0.2--4.0 keV energy band.
Worral et al. (1994) state that a combination
of a point source and of an extended emission, described by a $\beta$
model with 
core radius between $18^
{\prime\prime}$ and $65^{\prime\prime}$, provides a better fit to the
X--ray
data.

The angular size of the radio structure is 18 arcsec and
we have modeled the radio volume (ellipsoid semi--axis $62\times 22$ Kpc)
with reference to the 1.4 GHz map of 
Ronghui et 
al. (1992). 
If 3C 280 lies on the plane of the sky, the IC contribution to the soft
X--ray 
luminosity from the scattering of the 
CMB photons could be as high as 22$\%$ 
and the luminosity of a hidden 
quasar which would be required to IC produce  
the soft X--ray total luminosity would be about that of the {\it typical}
quasar, 
i.e. $L_{<Q>}$.
The total
angular size along the radio axis of the X--ray source predicted by our
model 
would be $\sim$ 12 arcsec for an e--folding drop in the luminosity
profile ({\bf Fig.1}). 
The ratio between the X--ray luminosities of the two lobes could range
from
1 to 4 depending on the inclination of the radio axis with respect to
the
line of sight. 
On the other hand
if, according to Worrall et al. (1994), it is assumed that the
point--like
source represents only 60$\%$ of the total luminosity and that the 
remaining 40$\%$ is due to cluster emission, then the 
required hidden quasar luminosity 
would be $\sim$ 0.52 $L_{<Q>}$, while the IC contribution due to the CMB
photons
would increase to $\sim 38\%$.
 Our model may represent a straightforward 
interpretation of the X--ray properties of this radio galaxy.

\subsection{{\bf 3C 294} (z=1.786)}

3C 294 is a well known example of the optical/radio alignment 
(McCarthy et al. 1990, McCarthy 1993).
The monochromatic luminosity of the detected 
Ly$\alpha$ line is $7.6 \times 10^{44}$
erg s$^{-1}$ and it shows large intrinsic widths (700--2600 km s$^{-1}$).
A possible interpretation of the observed properties 
has been suggested by McCarthy et al. (1990) 
assuming a powerful non-thermal central source 
with a luminosity comparable to that of a luminous quasar
which ionizes the gas clouds.
3C 294 has been observed with ROSAT PSPC (Crawford \& Fabian 1996b) for a 
total exposure time of 22.69 ksec.
The signal is very faint giving only $\sim 25$ source net counts.
The observed flux is consistent with either a thermal spectrum 
or an unabsorbed power law.
Assuming a power law spectrum with spectral index $\alpha$ =
1.1, consistent with the radio one, a 0.2--4.0 keV rest frame luminosity
of $\sim 3.0 \times 10^{44}$ erg s$^{-1}$ is
inferred.

The radio structure has been modeled (ellipsoid semi--axis $95\times 30$ 
Kpc) by making use of 5 GHz map of 
Strom et al. (1990). It has an angular size of 20 arcsec.
The contribution due to the IC scattering of the CMB photons is 
 $\sim 2.9 \times 10^{43}$ erg s$^{-1}$, i.e. $\sim$ 10\% of 
the soft X--ray luminosity.
The luminosity of the hidden quasar required to match the total soft 
X--ray luminosity
would be $\sim$ 3.6--4.5 $L_{<Q>}$ depending on the inclination
of the radio axis on the sky plane. This is a very luminous quasar, but 
we note that it may be consistent with the hypothetical
radiation source required to explain the observed strength of
the Ly$\alpha$ emission line. 
The predicted ratio between the X--ray luminosities of the two 
lobes would be between 1 and 7 depending on the inclination of 
the radio galaxy on the sky plane. In the case of a unit ratio the total
angular size of the 
X--ray source along the radio axis would be $\sim$ 10 arcsec.

\subsection{{\bf 3C 324} (z=1.206)}

This radio galaxy shows optical/radio alignment 
(McCarthy 1993) and a high degree
of polarization ($\sim$ 18\%) in the R--band 
(di Serego Alighieri et al. 1993).
From a recent modeling based on dust scattering of the nuclear light 
 the V magnitude of the hidden quasar
is estimated to be $m_V \sim 18.5$ (Manzini \& di Serego Alighieri 1996) .
3C 324 has been observed with the ROSAT PSPC (Crawford \& Fabian 1996b)
for a total exposure time of 15.4 ksec.
The spectral shape is ill defined being consistent with either a 
thermal spectrum or with an highly absorbed power law.
Assuming the power law spectrum predicted by our model from the radio data 
($\alpha$ = 1.07) we find a 
a rest--frame 0.2--4.0 keV luminosity $\sim 1.7 \times 10^{44}$
erg s$^{-1}$, a rather luminous X--ray source.

The radio structure has been modeled (ellipsoid semi--axis $58\times 23$ Kpc)
by making use of the 1400 MHz map from
 Fernini et al. (1993). Its extension subtends an angle of $\sim$ 14 arcsec.
The IC contribution due to the scattering of the CMB photons is
only $\sim$ 4\% and it does not change significantly with the inclination of the
radio axis with respect to the line of sight, while the luminosity of the
hidden quasar which would be required to match the total soft X--ray luminosity
is 5.1--6.0 $L_{<Q>}$ depending on the 
inclination of the radio axis. If one assumes 
the {\it typical} quasar spectrum of our model, the inferred 
V magnitude would be 16.2--16.4
at variance with the estimate of Manzini \& di Serego Alighieri 
reported above.
It should be noted that recent HST observations suggest that 3C 324
lies in a cluster of galaxies (Crawford \& Fabian 1996b and ref.
therein) whose thermal emission may provide a more satisfactory explanation
of this X--ray luminous source.

\begin{table*} 
\caption[]{The soft X--ray luminosity and source angular sizes of six powerful
FRII radio galaxies from the IC scattering of hidden quasar and CMB photons.}
\begin{center} \begin{tabular}{ccccccccc}
\hline
(1) & (2) & (3) &(4)&(5)&(6)&(7)&(8)&(9)\\
Name & z & $\theta_{ax}$ & $B_{eq}$ & $L_X(<Q>)$ & $L_X(CMB)$ & 
$L_Q/L_{<Q>}$ & $l$ & e-fold\\
  & & & (10$^{-5}$G)& (10$^{43}$ erg s$^{-1}$) &(10$^{43}$ erg s$^{-1}$)&
 & & (arcsec)\\ 
\hline

3C 277.2 & 0.766 & 90$^{\circ}$ & 3.17 & 2.28 & 0.56 & 0.93 & 1   & 10\\
         &       & 60$^{\circ}$ & 3.04 & 2.42 & 0.61 & 0.80 & 3.1 &   \\
         &       & 45$^{\circ}$ & 2.92 & 2.64 & 0.66 & 0.76 & 4.4 &   \\
3C 280   & 0.996 & 90$^{\circ}$ & 4.65 & 5.15 & 1.42 & 1.08 & 1   &  6\\
         &       & 60$^{\circ}$ & 4.47 & 5.48 & 1.53 & 0.99 & 2.8 &   \\
         &       & 45$^{\circ}$ & 4.24 & 5.61 & 1.67 & 0.94 & 3.8 &   \\
3C 294   & 1.786 & 90$^{\circ}$ &10.03 & 6.12 & 2.98 & 4.56 & 1   &  5\\
         &       & 60$^{\circ}$ & 9.66 & 6.89 & 3.22 & 4.01 & 4.2 &   \\
         &       & 45$^{\circ}$ & 9.23 & 7.44 & 3.54 & 3.68 & 7.1 &   \\
3C 324   & 1.206 & 90$^{\circ}$ & 9.37 & 4.73 & 1.12 & 6.10 & 1   &  3\\
         &       & 60$^{\circ}$ & 8.90 & 5.23 & 1.21 & 5.50 & 4.4 &   \\
         &       & 45$^{\circ}$ & 8.38 & 5.50 & 1.31 & 5.21 & 8.6 &   \\
3C 356   & 1.086 & 90$^{\circ}$ & 2.56 & 2.88 & 3.90 & 3.70 & 1   & 18\\
         &       & 60$^{\circ}$ & 2.43 & 3.15 & 4.21 & 3.25 & 2.8 &   \\
         &       & 45$^{\circ}$ & 2.29 & 3.32 & 4.61 & 2.97 & 3.6 &   \\
3C 368   & 1.132 & 90$^{\circ}$ & 9.71 & 8.01 & 0.54 & 1.85 & 1   &  4\\
         &       & 60$^{\circ}$ & 9.42 & 9.02 & 0.57 & 1.64 & 4.9 &   \\
         &       & 45$^{\circ}$ & 8.80 & 9.84 & 0.66 & 1.49 &10.0 &   \\
\hline
\end{tabular} 
\end{center}

{\bf Col.1} Source name;
{\bf Col.2} Redshift;
{\bf Col.3} Angle between the radio--axis and the line of sight;
{\bf Col.4} Values of the equipartition magnetic field in the lobes (Eq.A3);
{\bf Col.5} The IC X--ray luminosity in the 0.2--4.0 keV band (source frame)
assuming a {\it typical} hidden quasar;
{\bf Col.6} The IC X--ray luminosity in the 0.2--4.0 keV band (source frame)
from the scattering of the CMB photons;
{\bf Col.7} The required luminosity of the hidden quasar (in unit of a {\it
typical} quasar) to match the observed X--ray luminosity;
{\bf Col.8} The predicted X--ray ratio between the two lobes;
{\bf Col.9} The predicted angular size (radius) for an e--folding drop in the
luminosity profile.
 \end{table*}

\subsection{{\bf 3C 356} (z=1.086)}

3C 356 has been observed with the ROSAT 
PSPC for 18.6 ksec 
and detected at the $\sim 6\sigma$ level for a total of $\sim$ 30 net counts 
(Crawford \& Fabian 1993). While
the spectrum is well fitted by a
hot thermal plasma model, it is inconsistent with an 
absorbed power law, Cygnus A type.
It should be stressed, however, that an unabsorbed power 
law with spectral index $\sim$ 1.0, such as that predicted by our IC model, 
gives an adequate representation of the data. By adopting this spectral slope
one derives a luminosity $\sim 1.4 \times 10^{44}$ erg s$^{-1}$ in the 0.2--4.0
keV energy interval.
From the analysis of the PSPC data it has been suggested that the source is
extended, a good description being that of a point--like 
 source (with radius $r< 12^
{\prime\prime}$) contributing $\sim 20\%$ of the flux and an extended component 
approximately aligned with the radio axis. This has been confirmed by a
subsequent observation with the ROSAT HRI, which failed to detect a point source
at the position of the radio galaxy at the 
flux level expected from the PSPC observation 
(Crawford \& Fabian 1996a). These authors suggest that the most likely 
interpretation of the source is that of the emission from a hot intracluster 
gas in a cluster of galaxies hosting 3C 356.

The radio structure (ellipsoid semi--axis $290\times 70$ Kpc)
has been derived from the 1490 MHz map of Leahy et al. (1989). The angular
size subtended by the hot spots is $\sim$ 80 arcsec.
Since 3C 356 is the radio galaxy with the largest volume in our 
sample,
 the IC contribution from the scattering of the CMB photons is expected
to be important. Assuming that the radio galaxy 
lies on the plane of the sky, we find 
that this contribution accounts for $\sim$ 37\% of the soft X--ray luminosity.
The inferred hidden quasar luminosity, which would be required to account for 
the remaining soft X--ray luminosity of the source,
is $\sim$ 3.7 $L_{<Q>}$. The predicted angular size of the source along 
the radio axis would be $\sim$ 36 arcsec for an e--folding drop in the radial
luminosity profile.
If the inclination of the radio axis with respect to the line of sight is
substantial, say $\theta_{ax} = 60^{o}$, the efficiency of the IC model increases
and we expect an asymmetric source with a luminosity ratio $\sim$ 3 between 
the two X--ray lobes. 

New optical spectropolarimetric data (Cimatti et al. 1997) obtained with
the Keck Telescope show that the two radio--optical components (a \& b) are 
polarized in the UV rest frame continuum emission and that there is a broad
MgII$\lambda$2800 emission line both in total and polarized light associated
with component (a), indicating that 80\% of the 2800$\AA$ flux is contributed 
by non--stellar radiation. These authors conclude that their observations
definitely support the AGN unification model, but that 
it remains unclear which
of the two components is hosting the hidden quasar: if it is located in 
component (b), then its luminosity would have to be at least as high as the  
maximum luminosity observed in 3C quasars, while if located in component (a) 
the energy requirements would be greatly alleviated leading to a luminosity
closer to that of a {\it typical} quasar. The luminosity of the hidden quasar
required by our model is consistent with these findings.  

\subsection{{\bf 3C 368} (z=1.132)}

This is one of the best studied z $\sim$ 1 radio galaxies.
It shows an imponent emission region oriented along the radio axis (Chambers 
et al. 1988).
Several models have been proposed to interpret the
emission lines and their complex velocity structure and 
the elongated continuum morphology.
Djorgovski et al. (1987) suggest an interpretation in terms of 
a violent star burst caused by a merger between galaxies, while
Meisenheimer \& Hipplelein 
(1992) suggest that both the emission lines luminosity 
and their 
spatial and kinematic structure can be explained by the interaction between 
the radio jets and a high density ($n_e \sim 0.1$ cm$^{-3}$) surrounding medium.
However, both these models could not account for the detection of extended 
strong optical polarization, a {\it prima facie} evidence of scattered light
from a hidden, luminous quasar
(Cimatti et al. 1993 and ref. therein). More recent investigations, although 
not inconsistent with the possible existence of a hidden quasar, failed 
detecting any polarized component (upper limits of 2--3\%) in 3C 368,
while at the same time they lend further support to the idea that the strong
emission associated with the aligned material is due to the interaction
with the radio jets (Stockton et al. 1996 and ref. therein).
3C 368 has been observed with the ROSAT PSPC (Crawford \& Fabian 1995) for 
a total exposure time of 26.4 ksec. It was detected at the $\sim$ 5$\sigma$ 
level giving only $\sim$ 20 source net counts.
Assuming a power law spectrum with $\alpha$ =  
1.1, consistent with the radio spectral index, we find that the 0.2--4.0 keV 
rest frame luminosity is
$\sim 1.5 \times 10^{44}$ erg s$^{-1}$.

The model radio structure (ellipsoid semi--axis $54\times 18$ Kpc)
has been derived from the 5 GHz map of 
Chambers et al. (1988). 
The radio source is rather small, subtending an angle of
only 14 arcsec, which means that the associated rather strong 
equipartition field
prevents any major contribution ($\sim$ 4\%) from the IC scattering of CMB 
photons.
The required luminosity of the hidden quasar to IC produce the 
observed soft X--ray luminosity is $\sim$ 1.5--1.9 $L_{<Q>}$,
depending on the inclination of the radio axis. The predicted angular size of 
the X--ray source would be $\sim$ 4 arcsec.
Assuming an inclination of the radio axis $\theta_{ax}$ = 60$^{o}$, 
the required hidden quasar luminosity would be $\sim$ 1.6 $L_{<Q>}$ and 
the luminosity ratio between the two X--ray lobes 
 $\sim 5$. We conclude that the soft X--ray luminosity of 3C 368 may
be entirely accounted for by our model.

\subsection{\bf The case of Cygnus A} 

One may wonder whether our model could be tested in the case of the (close by)
archetypal FRII radio galaxy Cygnus A.
Assuming the presence of a {\it typical} quasar in the nucleus of Cygnus A, 
we find an IC luminosity in the 0.2 -- 4.0 keV band  
$\sim 2 \times 10^{43}$ erg s$^{-1}$, which is
$\sim$ 50--100 times lower than the thermal emission
of the surrounding hot intracluster gas.
Furthermore, the  predicted e-folding drop in the luminosity profile
corresponds to an angular size 
($\sim$ 30 arcsec)  of the same order as that of the cluster emission.

Even assuming that the brightness distribution predicted by
our model (roughly $\propto 1/r$) could be extrapolated well inside
the central region, its value would still be small (at most 10-15 \%) compared
to that of the cluster emission.
One can conclude that the soft X-ray luminosity predicted by our model,
although of similar strength as those derived for the high redshift 
powerful FRII radio galaxies, is shadowed by the presence of the
strong cluster emission.
ROSAT HRI observations of the central part of Cygnus A have revealed a 
point-like source, identified with the unabsorbed quasar emission, and
the possible presence of a weak extended emission
($\sim$ 10 arcsec; Harris et al. 1994b) in addition to that of the
cluster gas. If confirmed, such extended emission could be
consistent with our model.

\section{\bf Summary and Conclusions}

We have shown that, in the framework of the unification scheme relating
quasars and FRII radio galaxies, the contribution
to the soft X-ray emission of strong radio galaxies via the IC scattering
of the optical-IR 
photons from a hidden quasar may be important.
Given a {\it typical} quasar spectrum, the main contribution comes from 
the IC scattering of the IR photons under the assumption that the 
spectrum of the radio emitting electrons can be extrapolated down to lower
energies ($\gamma \sim 100-200$). 
For all sources discussed in this work 
our estimates of the
particles and fields energy densities are 
derived from minimum energy condition
by imposing the same low energy cut off  ($\gamma_{min}$ = 20) in the particle
distribution.
There is no direct observational evidence of how far one can extrapolate
the radio electron spectrum to lower energies. The assumption $\gamma_{min}=20$
appears consistent with present knowledge of the physical conditions
prevailing in the radio sources. If we had  chosen a larger $\gamma_{min}$,
but less than $\sim$ 100, then we would have obtained larger normalizations
of the electron spectra and larger predicted soft X--ray luminosities.
This statement holds even taking into account the fact that the contribution
from the IC scattering of the optical--near IR photons emitted
by a hidden quasar would progressively tend to zero.
In this case , however, the predicted soft X--ray spectrum would be 
significantly flatter.
 
Given the directionality of the photon flux propagating outward from a  
hidden quasar, the standard formulae for the computation of the 
IC emission are not applicable. We have derived 
the general formulae for the anisotropic IC scattering in the Thompson 
approximation and discussed them in 
the Appendix B together with those specifically applied to our model.

The observed soft X-ray properties of six strong FRII radio galaxies have been
compared with those predicted by our model.
Such objects are some of the stronger high redshift 
FRII radio galaxies in the 3CR sample for which X-ray
data are available at the moment. Our estimates include the contributions to the
soft X--ray fluxes due to the IC scattering of the radio electrons with the
CMB photons which, because of the high redshifts involved, may be significant.
In our computations we have assumed that the radio volumes can be approximated
by prolate ellipsoids filled with a uniform distribution of particles and 
fields and that the energy density of the relativistic electrons equals that
of the protons.
These volumes  may be significantly larger then those filled
by the relativistic particles, but in Sect. 2.3 we have shown that 
this may lead to larger predicted values only for the fraction (generally
small) of the total X--ray luminosity from the IC of the CMB
photons.

We find that for three radio galaxies (3C 277.2, 3C 280 and
3C 368) the model predicted soft X-ray luminosities agree with those
observed if a quasar of average strength is supposed to be hidden in 
their nuclei. It is interesting to note that
in the case of 3C 277.2 the derived optical luminosity of the
hidden quasar is in very good agreement with that inferred from 
spectropolarimetric studies. The predicted soft X--ray 
spectrum of 3C 280 is compatible 
with that observed  and with the evidence of a low $N_H$ column
density. 
For the remaining three radio galaxies of our sample 
(3C 294, 3C 324 and 3C 356) we find that the hidden quasars must be
significantly more powerful than the {\it typical} quasar in order
to match the observed soft X-ray luminosities.
In the case of 3C 294 the hidden quasar would have to be $\sim$ 1.5 magnitude
more luminous than the {\it typical} quasar, which may be consistent with
the very strong Ly$\alpha$ emission observed in this radio galaxy.
A similarly powerful hidden quasar is required for 3C 356, in which case 
our model may provide a possible interpretation of both the
extension and alignment with the radio structure of the soft X--ray source 
and of its (poorly constrained) spectral shape. The presence of a luminous
hidden quasar is strongly supported by recent spectropolarimetric studies
of this source.
In the case of 3C 324 the required optical-IR luminosity of the hidden
quasar is about 2 magnitudes brighter than that of the {\it typical} quasar
and considerably higher than that inferred from spectropolarimetric
studies, but recent HST observations suggest
that 3C 324 may lie in a cluster of galaxies 
whose thermal emission is likely to dominate the X--ray flux detected in 
the direction of this radio galaxy.
In addition, we have briefly discussed
the special case of Cygnus A
showing that the soft X--ray luminosity predicted by our model is completely
dwarfed by the very strong emission of the surrounding hot intracluster gas.
Thanks to the solution of the anisotropic IC equations, for each radio galaxy
we are able to 
compute the expected ratio between the X-ray luminosities of the lobes 
as a function
of the inclination of the radio axis on the sky plane and of the differential
electron spectrum. 
This ratio may be large and it can provide
a direct observational handle on the applicability of our model. If detected, 
this effect could also tell us which of the two lobes is the farthermost.
Unfortunately
 our model forecasts an extended X-ray emission
 with a somewhat steep brightness profile
(roughly $\propto 1/r$) so that the small 
($<$ 20 arc sec) angular size of 
the distant strong FRII radio galaxies and their relatively weak X-ray
emission make these observational tests difficult to be performed with present 
X-ray facilities.
The expected X--ray luminosity profile ({\bf Fig.1})
shows that a depletion of the relativistic particles in the
mid--central source volume may significantly decrease the efficiency of our 
model.
Under the conservative assumption whereby the  relativistic electrons 
are only present within the radio lobes defined by the high frequency radio 
maps, we find that the estimates of the X--ray luminosities of Table 1 are
reduced by factors from 2 to 3.

Clearly the predictions of our model are based on various assumptions
concerning the physical properties of the FRII radio galaxies.
We believe to have shown that reasonable choices of the parameters lead to
results that can be  tested with observations.
In particular, the imaging and spectral capabilities of the future
X--ray facilities, such as AXAF, may provide a unique tool to study
the distribution and spectra of the relativistic electrons at lower
energies thus bringing an important information on the structure
and evolution of these sources.

\begin{acknowledgements}
The authors would like to thank A. Cimatti, S. di Serego Alighieri, 
A.C. Fabian,
R. Fanti and G. Ghisellini
for useful discussions and informations
and the anonymous referee whose constructive criticism and suggestions
have considerably improved the presentation of this work. 
AC acknowledges partial financial support from the 
Italian Space Agency under the contract ASI-95-RS-152.
This research has made use of the NASA/IPAC Extragalactic Database (NED)
which is operated by the Jet Propulsion Laboratory, Caltech, under contract
with the National Aeronautics and Space Administration.
\end{acknowledgements}

\appendix
\section{\bf Equipartition equations}

In this Appendix we derive  equipartition quantities based on a low
energy cut-off in the particle energy distribution, instead of a low frequency
cut-off in the emitted synchrotron spectrum.
Under the assumption that the differential energy
distribution of the relativistic electrons is represented by a single power
law, $N(\gamma)=K_{e} \gamma^{-\delta}$, the total
(particles and fields) energy density 
can be written as:

\begin{equation}
w_{tot} = {{B^2}\over {8 \pi}}+ (1+F)\,mc^2 
\int_{\gamma_{min}}^{\gamma_{max}} {K_e \gamma^{-\delta+1} d\gamma}
\end{equation}

where $F$ is the ratio  between the energy densities of the relativistic
proton and electron components and  $m$ is the electron mass.
From the
standard synchrotron theory 
 $K_e= j(\nu) a(\delta) B^
{-{{\delta+1}\over 2}} \nu^{{\delta-1}\over 2}$, 
where  $j(\nu)$ is the synchrotron  emissivity
at a frequency $\nu$ and 
$a(\delta)$ a constant.
Assuming $\delta >2$  and $\gamma_{max} >> \gamma_{min}$,
Eq.(A1) becomes:

\begin{equation}
w_{tot}= {{B^2}\over {8 \pi}}+{{(1+F) a(\delta)\nu^{{\delta-1}\over 2}
\,mc^2 }\over{\delta-2}} 
B^{-{{\delta+1}\over 2}}j(\nu)\gamma_{min}^{2-\delta} 
\end{equation}

By imposing the
minimum energy condition,
under the assumption that the particle and magnetic field energy densities
are uniformly distributed in a volume $V$, and that
particle momenta
and  field lines are randomly distributed, one finds the equipartition
magnetic field strength

\begin{equation}
B_{eq}=\left[C(\delta)
 {{P(\nu)}\over{V}}
\nu^{{{\delta-1}\over2}} (1+F)\right]^{2\over{\delta +5}}
\gamma_{min}^{{2(2-\delta)}\over (\delta +5)} 
\end{equation}

with

\begin{eqnarray*}
C(\delta) =
{{ \Gamma\left({{\delta}\over 4}
+{7\over 4}\right)}\over{\Gamma\left({{\delta}\over4} 
+{{19}\over{12}}\right) \Gamma\left({{
\delta}\over4}-
{1\over{12}}\right) \Gamma\left({{\delta}\over 4}+{5\over 4} \right)}}
\left({{2\pi mc}\over{3e}}\right)^{{\delta-1}\over2}
\cdot\nonumber\\
4 \sqrt{\pi/ 3} e^{-3} (mc^2)^2
(\delta +1)^2 (\delta -2)^{-1}
\end{eqnarray*}

where $P(\nu)$ is the source synchrotron power at a frequency $\nu$,
$\Gamma$ is the Euler function, and  $e$ 
the (positive) electron  charge.
Representative values of $C(\delta)$ are given in Table 2.
It should be noticed that, at variance with the classical
equipartition equations, here the energy density of the relativistic
particles
is exactly equal to that of the magnetic field 
(instead of 4/3).

It may be helpful for practical applications
to write down the relationship between 
our equipartition
quantities and those (primed) 
derived from the standard equipartition equations with
reference to the observed radio flux densities in the 10 MHz -- 100 GHz
frequency band (source frame). 
We find:

\begin{equation}
B_{eq} = D(\delta)
\gamma_{min}^{{2(2-\delta)}\over{\delta+5}}
 (B^{\prime}_{eq})^{{7}\over{\delta+5}}
\end{equation}

where $D(\delta)$ is $\sim 1$ (Table 2).
It should be noticed that $B_{eq} > B^{\prime}_{eq}$ for $B^{\prime}_{eq}
< D(\delta) \gamma_{min}^{-2}$, in which case
$K_{e} < K^{\prime}_{e}$ follows from the synchrotron emissivity formulae
(i.e. $K_e B_{eq}^{(\delta+1)/ 2}=
K_e^{\prime} (B_{eq}^{\prime})^{(\delta+1)/ 2}$).

\begin{table}[h] 
\caption[]{Numerical values of $C(\delta)$ and $D(\delta)$ for some values of
$\delta$}
\begin{center}
\begin{tabular}{ccc}
\hline
$\delta$ & $C(\delta)$ & $D(\delta)$\\
\hline
2.3 & 8.87 E12& 1.00\\
2.5 & 2.76 E12& 1.01\\
2.7 & 4.75 E11& 1.05\\
2.9 & 8.95 E10& 1.09\\
3.1 & 1.73 E10& 1.14\\
3.3 & 3.37 E9& 1.17\\
\hline
\end{tabular} \end{center} \end{table}

\section{\bf Anisotropic inverse Compton scattering}

In this Appendix we derive the basic formulae for the computation
of the X-ray emission predicted by our model as a result of the
IC scattering of the photons from a hidden quasar with the 
relativistic electrons in the radio lobes.
We note that such a calculation can not be performed with
the standard IC equations usually found in the literature (Jones 1968; 
Blumenthal \& Gould 1970) essentially for two main reasons:
first, in our quasar model the  relativistic electrons
are flooded by an anisotropic photon flux and, second,
the ultra--relativistic limit ($\beta = 1$) may not be completely applicable.
The computation has been performed in three successive steps.
First of all the IC spectrum resulting from the interaction
between an unidirectional photon beam and an electron of energy
$mc^2\gamma$ has been derived.
Then the IC emissivity due to the interaction 
between an unidirectional photon beam and a population of relativistic 
electrons with a power-law energy distribution and 
an isotropic distribution of their momenta has been computed.
The anisotropic emission is described as a function of the angle 
between the photon beam and the line of sight.
Finally the total IC emission toward the observer is obtained by 
integrating the emissivity over the
radio lobe volume for a given inclination of the
radio galaxy on the sky plane and for a given angular distribution
of the radiation from the hidden quasar.

\noindent
All the calculations have been performed in the Thompson limit.

\subsection{\bf Scattering of a photon beam by a single electron.}

First of all we define the geometry of the scattering process by
assuming that the motion of the incident photons is along the z-axis ({\bf
Fig.2}).

\begin{figure}
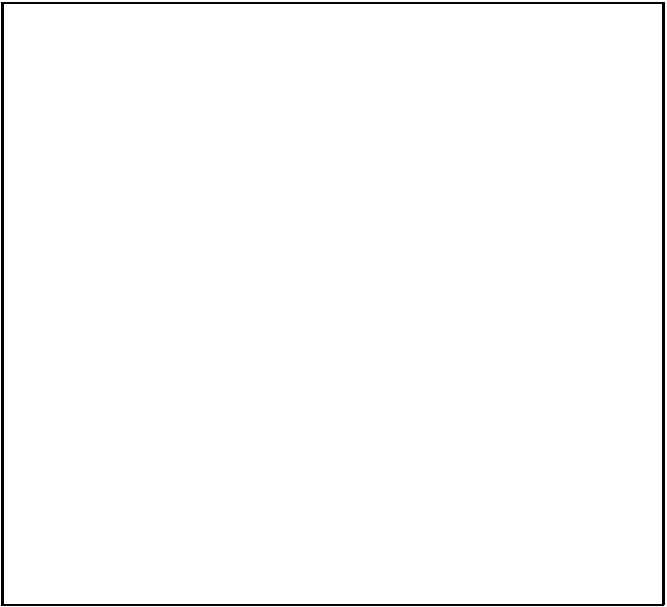

\picplace{8cm}
\caption[]{The geometry of the scattering between one electron
with momentum in the ($\theta$, $\phi$) direction and a photon beam 
propagating along the z--axis. 
The scattered photon direction is indicated by the ($\theta_{SC}$, $\phi_{SC}$)
coordinates.}
\end{figure}

\noindent
The momenta of the incident 
and scattered photons  (of energy $\epsilon$ and $\epsilon_1$)
and of the  electron  are 
respectively described by the four-vectors:

\begin{equation}
P={\epsilon \over{c}} \left(1,0,0,1\right) 
\end{equation}

\begin{equation}
P_1={\epsilon_1 \over{c}} \left(1,k_1,k_2,k_3\right) 
\end{equation}

\begin{equation}
P_{e}=mc\gamma \left(1,\beta e_1,\beta e_2,\beta e_3\right)
\end{equation}

\noindent
where $k_j$ and $e_j$ represent the unit vector components
in a spherical coordinate system and $\beta \equiv v/c$.

Let us consider a monochromatic unidirectional photon
beam (of energy $\epsilon_i$)
so that the photon number density seen by an electron per unit 
solid angle and energy is

\begin{equation}
n(\epsilon,x,\phi)d\epsilon dx d\phi=
n\, \delta(\epsilon-\epsilon_i) \delta(x-x_i) \delta(\phi-\phi_i)
d\epsilon dx d\phi
\end{equation}

where $\theta$ and $\phi$ 
are the angles that describe 
the electron
momentum in spherical coordinates, 
$x=\cos\theta$ and $\delta$ is the $\delta$--function.

The ratio $dn/\epsilon$ is known to be a relativistic invariant 
(e.g. Blumenthal \& Gould 1970), so that from (B4) the total differential 
photon density in the electron frame (primed) is:

\begin{equation}
{{dn^{\prime}(\epsilon^{\prime},\Omega^{\prime}
;\epsilon,\Omega)}\over{d\epsilon
^{\prime} d\Omega^
{\prime}}}= n\, \delta( \epsilon- \epsilon_i){{d\Omega d\epsilon }\over
{d\Omega^{\prime} d\epsilon^{\prime}}}
{{\epsilon^{\prime}}\over{\epsilon}} \delta(x-x_i) \delta(\phi- \phi_i)
\end{equation}

The angular and energy distribution of the scattered photons per unit time
in the electron
frame is:

\begin{equation}
{{dN_{\gamma,\epsilon}}\over{dt^{\prime} d\epsilon^{\prime}
d\Omega^{\prime}_{SC} d\Omega^{\prime} d\epsilon^{\prime}_1}}=
{{dn^{\prime}(\epsilon^{\prime},\Omega^{\prime};\epsilon,\Omega)}
\over{d\epsilon^{\prime} d\Omega^{\prime}}}
{{c \,d\sigma}\over{d\Omega^{\prime}_{SC} d\epsilon^{\prime}_1}} 
\end{equation}

where $N$ = the total number of scatterings (invariant), $c$ is 
the light velocity and

\begin{equation}
{{d\sigma}\over{d\Omega^{\prime}_{SC} d\epsilon^{\prime}_1}}=
{{r_0^2}\over 2} (1+\cos^2\theta^{\prime}_{SC}) \delta(\epsilon^{\prime}_1-
\epsilon^{\prime}) 
\end{equation}

is the differential Thompson cross section, $\theta^{\prime}_{SC}$ being the 
angle between the incident
and scattered photon in the electron frame, $d\Omega^{\prime}_{SC}
=-d(\cos\theta^{\prime}_{SC}) \, d\phi^{\prime}_{SC}$ and $r_0$ 
the classical electron radius.

The energy distribution of the scattered photons in the electron frame is thus:

\begin{eqnarray} 
{{dN_{\gamma,\epsilon}} \over {dt^{\prime} d\Omega^{\prime}_{SC} 
d\epsilon^{\prime}_1}}=
{{r_0^2 c}\over 2}  \int \int_{(\epsilon^{\prime},\Omega^{\prime})}
{{dn^{\prime}(\epsilon^{\prime},\Omega^{\prime};\epsilon,\Omega)}\over
{d\epsilon^{\prime} d\Omega^{\prime}}}\cdot \nonumber\\
(1+\cos^2 \theta^{\prime}_{SC}) \delta (\epsilon^{\prime}
_1- \epsilon^{\prime}) d\epsilon^{\prime} d\Omega^{\prime} 
\end{eqnarray}

From a Lorentz boost of Eqs.(B1),(B2),(B3) one obtains:

\begin{equation}
\cos\theta^{\prime}_{SC}=1+ {{(k_3 -1)}\over {\gamma^2 L
L_1}}
\end{equation}

\noindent
and

\begin{eqnarray}
 \cos(\phi^{\prime}_{SC}-\phi_0)=
\{\cos \theta^{\prime}_{SC} L_1 \Delta  
- L ( (\gamma-1)e_jk_je_3 +\nonumber\\  
k_3-\gamma \beta e_3 )\} 
\left[
\left( 1-\cos^2\theta^{\prime}_{SC}\right) \left(
1-\left(\Delta/L\right)^2\right)LL_1\right]^{-1}
\end{eqnarray}

\noindent
where 
$L=(1-\beta e_3)$,
$L_1= (1-\beta e_j k_j)$, $\Delta=(\gamma-1)e_3^2 +1 -\gamma\beta e_3$ and 
$\phi_0$ an arbitrary phase.
The  Jacobian of $\Omega^{\prime}_{SC}$ with respect to 
$\Omega_{SC}$ is: 

\begin{equation}
{{d\Omega^{\prime}_{SC}} \over {d\Omega_{SC}}}={1\over{
\gamma^2 L_1^2}} 
\end{equation}

Finally, 
the emitted power spectrum per unit frequency and solid angle 
in the lab frame is obtained by considering the 
relativistic transformation $\gamma dt^{\prime}=dt$,
$d\epsilon^{\prime}_1=\gamma (1-\beta e_jk_j) d\epsilon_1$, 
$\epsilon^{\prime}=\gamma \epsilon (1-\beta e_3)$ and Eqs.
(B5), (B8), (B9), (B11).
We obtain:

\begin{eqnarray}
 {{dE(P_{e},P,P_1)}\over{dt d\epsilon_1 d\Omega_{SC}}}= 
{{c r_0^2}\over {2\gamma^2}}
n\, {{\epsilon_1}\over {L_1}} 
\delta (\epsilon- \epsilon_i)
\left[\left({{(k_3-1)}\over{\gamma^2
LL_1}}\right)^2 \right.\nonumber\\
\left.
+2+2{{(k_3-1)}\over{\gamma^2LL_1}}\right] 
\end{eqnarray}

\noindent
together with the additional equation (with $\epsilon_1 > \epsilon$)

\begin{equation}
e_jk_j  ={1\over {\beta}}\left[1-{{\epsilon}\over{\epsilon_1}}
L\right]
\end{equation}

\subsection{\bf Scattering of a photon beam by electrons with isotropic 
momenta distribution}

The  emitted power per unit frequency and solid angle is obtained 
by integrating Eq.(B12) over
$\gamma$, $e_3$, and $\phi$
assuming an isotropic distribution of the electrons momenta
and a power law energy differential spectrum
$ N_{e}(\gamma) =
{1\over {4\pi}} K_e \gamma^{-\delta}$

\begin{eqnarray}
{{dW(\epsilon,P_1)}\over{dt d\epsilon_1 d\Omega_{SC}}}=
{{c r_0^2 K_e}\over{8\pi}} n \int \int \int
\gamma^{-(\delta+2)} \epsilon_1^2 
\delta(\epsilon- \epsilon_i)\cdot
  \nonumber\\
{1\over {L \epsilon}}
\left[2+\left({{(k_3-1)}\over{\gamma^2 L^2 }} {{\epsilon_1}\over 
{\epsilon}}\right)^2
+2{{(k_3-1)}\over{\gamma^2L^2}} {{\epsilon_1}\over 
{\epsilon}}  \right] de_3d\phi d\gamma 
\end{eqnarray}

\noindent
From Eq.(B13) and the $\delta$-function properties we find:

\begin{eqnarray}
\delta (\epsilon - \epsilon_i)=
\delta(\gamma-\tilde{\gamma}) \gamma^3 \beta \epsilon_1 
{{(e_3 -e_jk_j)}\over {(\epsilon_1 e_jk_j -\epsilon e_3)^2}}=\nonumber\\
=\delta(\gamma -\tilde{\gamma}) {{\gamma^3 \beta}\over{ \epsilon_1}}
{{L^2}\over 
{\left[e_3\, (1-k_3) -e_1k_1 -e_2k_2 \right]}}
\end{eqnarray}

\noindent
where

\begin{equation}
\tilde {\gamma} = 
\left[ 1- \left({{\epsilon_1 - \epsilon_i}\over {\epsilon_1 e_jk_j - 
\epsilon_i e_3}} \right)^2 \right]^{- {1\over 2}}=\left[1-\tilde{\beta}^2
\right]^{-1/2}
\end{equation}

\noindent
is the Lorentz factor of the electrons for a particular scattering
configuration, i.e. having fixed $\epsilon_1$, $e_3$ and $e_jk_j$ for
a given $\epsilon_i$, with the condition $\beta e_jk_j \neq 1$.

\noindent
Due to the particular choice of the geometry, the problem is 
symmetric with respect to the z--axis.
For this reason we may assume that the
line of sight lies on the y--z plane and  
as a result, one has $k_2 =0$ and $k_1 = \sqrt{1-k_3^2}$.
From Eqs.(B15) and (B16), by renaming $\epsilon_i$ with $\epsilon$ 
and because $e_1= \sin \phi
\sqrt{1-e_3^2}$, 
we can then integrate Eq.(B14) over the energy of the electrons  to find

\begin{eqnarray}
{{dW(\epsilon,k_3)}\over{dt d\epsilon_1 d\Omega_{SC}}}=
{{c r_0^2 K_e n }\over {8 \pi}}  
{{\epsilon_1^2(\epsilon_1 - \epsilon)}\over{\epsilon}}
\int \int 
a(e_3,\phi)\cdot\nonumber\\
\left[2+({{(k_3-1)}\over {b(e_3,\phi)}})^2
+2{{(k_3-1)}\over {b(e_3,\phi)}} \right] 
de_3 d\phi
\end{eqnarray}

\noindent
where 

\begin{eqnarray}
& a(e_3,\phi)= \{ - (\epsilon_1  -\epsilon)^2+ 
& \nonumber\\
& [\epsilon_1
(\sin\phi\sqrt{1-e_3^2}\sqrt{1-k_3^2}+ 
e_3k_3)-e_3\epsilon]^2 
 \}^{{\delta-1}\over 2}\cdot&  \nonumber\\
&\left(\epsilon_1
(\sin\phi\sqrt{1-e_3^2}\sqrt{1-k_3^2} + e_3k_3)-  e_3\epsilon \right)^
{-(\delta +1)}&
\end{eqnarray}

\noindent
and

\begin{eqnarray}
&b(e_3,\phi)=\epsilon_1 \epsilon
\left[\sin\phi\sqrt{1-e_3^2}\sqrt{1-k_3^2} +e_3(k_3-1)\right]^2 \cdot &
\nonumber\\
&\{\left[\epsilon_1(\sin\phi\sqrt{1-e_3^2}\sqrt{1-k_3^2}
+e_3k_3)- e_3 \epsilon\right]^2& \nonumber\\
&-\left(\epsilon_1-\epsilon\right)^2\}^{-1}&
\end{eqnarray}

\noindent
Since Eq.(B17) cannot be integrated analytically, except for
particular values of the parameters, we have performed a numerical integration
over all possible 
scattering configurations.
The limits of integration are  obtained by requiring
that $0< \tilde{\beta} \leq 1$. The extremes of 
integration of $e_3$ are derived from Eq.(B16) by 
setting $\tilde{\beta} =1$:  

\begin{eqnarray}
e_3(\pm)=(1-\cos^2 \phi)^{1\over 2} 
\left[ {{(1-\epsilon / \epsilon_1)(k_3-\epsilon /
\epsilon_1)}\over{(1-k_3^2)^{1\over 2} (1-cos^2 \phi)^{1\over 2}}}\pm
\right. \nonumber\\
\left. \left(
(1-k_3^2)\,\sin^2 \phi +(k_3-{{\epsilon}\over{\epsilon_1}})^2
-(1-{{\epsilon}\over{\epsilon_1}})^2 \right)^{1\over 2}\right]
\cdot \nonumber\\
(1-k_3^2)^{1\over 2} \{(1-k_3^2)\sin^2 \phi +(k_3-{{\epsilon}\over
{\epsilon_1}})^2 \}^{-1} 
\end{eqnarray}
 
\noindent
The coincidence $e_3(-)=e_3(+)$ is obtained for a $\phi^{\ast}$ such that

\begin{eqnarray*}
\sin\phi^{\ast}=
\left(1- {{2\epsilon}\over {\epsilon_1 (1+k_3)}}
\right)^{1/2}
\end{eqnarray*}

\noindent
Then the scattering domains can be described as follows:

\noindent
for a line of sight such that
$k_3 >-1+2\epsilon / \epsilon_1$ (but $k_3 \neq 1$), 
then $\phi^{\ast}\, <\phi \, <\pi - \phi^{\ast}$
and $e_3(-) < e_3 < e_3(+)$; while for a line of sight such that
$k_3 \leq -1+2 \epsilon /\epsilon_1$,
one has either $0 \leq \phi < \pi$ and $-1 < e_3 < e_3(+)$ or
$\pi \leq \phi \leq 2\pi$ and $-1 < e_3 < e_3(-)$.

The (normalized) emitted power per unit 
frequency and solid angle is shown in {\bf Fig.3}
for three values of $\delta$ as a function of $\theta_{SC}$.
\begin{figure}
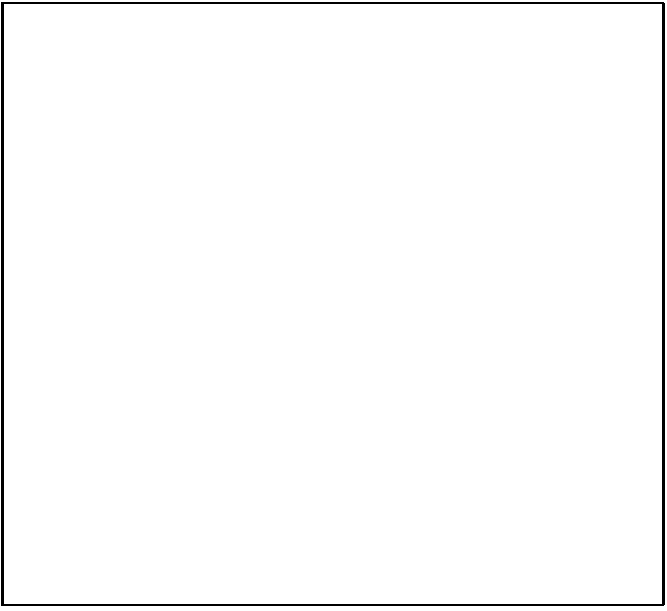

\picplace{8cm}
\caption[]{The normalized IC emitted power per unit frequency and
solid angle as a function of the scattering angle is plotted for three 
representative values of the electrons spectral index $\delta$: 3.0 
(solid line), 2.5 (dotted line) 
and 2.0 (dashed line).
}
\end{figure}

It should be noted that
the spectral slope depends on both the ratio $\epsilon_1/ \epsilon$
and $k_3$.
This is due to the varying geometry of the
scatterings and to the fact 
that we have relaxed the ultra--relativistic ($\beta=1$) approximation.
For example, assuming an energy of the incident photons $\epsilon =$ 0.1 eV 
and $\delta =2.5$, we find
that the spectral index between 10 and 20 eV is 0.727, 
0.737 and
0.754  respectively for $\theta_{SC} =$ 3.12, 2.20 and 1.50.
However, these differences  vanish  at somewhat higher 
energies, as it should be: for photon energies larger than $\epsilon_1 =$ 2 keV 
we recover the classical $\alpha$ =
0.750  for all the $\theta_{SC}$ 
(a more detailed discussion will be 
given elsewhere). From the standpoint of the model discussion presented
in this paper these differences can be neglected.

\subsection{\bf Application of the IC formulae to our source model}

Here the photon number depends on the distance $r$ from the
hidden quasar and it
can be easily computed for
a band (of lower and upper energy $\epsilon_L, \epsilon_U$) 
of the quasar spectrum.
If $L_{Q,b}$ is the quasar luminosity in such band and $d$ 
the spectral index, the photon number per unit volume and unit energy interval
is

\begin{equation}
n(\epsilon,r) = {{(d-1) \epsilon^{-(d+1)}}\over
{4\pi c r^2 \left[ \epsilon_{L}^{-(d-1)} - \epsilon_{U}^{-(d-1)}
\right]}} L_{Q,b} 
\end{equation}

\noindent
From Eqs.(B17) and (B21) one can write:

\begin{eqnarray}
{{dW(r,k_3)}\over{dt d\epsilon_1 d\Omega_{SC}}}=
{{r_0^2 K_e (d-1) \epsilon_1^2 }\over
{32 \pi^2 [\epsilon_{L}^{-(d-1)} - \epsilon_{U}^{-(d-1)}]}} r^{-2} L_{Q,b}
\cdot\nonumber\\
\int I(\epsilon_1,k_3;\epsilon,e_3,\phi)
\epsilon^{-(d+2)} d\epsilon
\end{eqnarray}

\noindent
where $I(\epsilon_1,k_3;\epsilon,e_3,\phi)$ is the integral in Eq.(B17)
with $n= n(\epsilon) d\epsilon$.

\begin{figure}
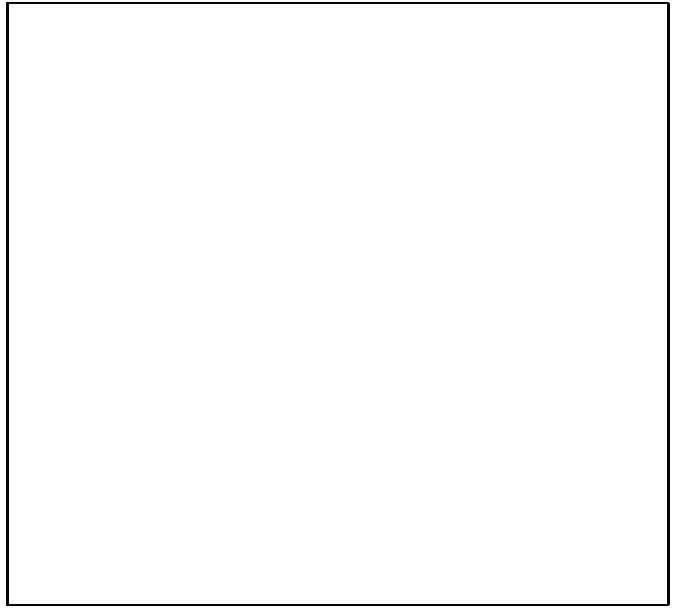

\picplace{8cm}
\caption[]{The ellipsoidal distribution of the relativistic
particles, for a given inclination angle ($\theta_{ax}$) of the radio axis 

with the line of sight, and the quasar emission cone.
The surface of the cone coaxial with the observer's direction identifies 
radiation paths with the same scattering configuration. The intersection
of this conical surface with the particle ellipsoid, within the quasar 
emission cone, at a distance $r$ 
from the nucleus defines the differential volume element 
$D(r,k_3)$ (thick curved line).}
\end{figure}

We assume that the radiation from the hidden quasar is confined
within a cone of half opening angle $\theta_Q$. Since it has been
shown that the observed
IC emission depends on the angle between the line of sight and the incident
photon beam,
in order to carry out the model computations it is convenient to introduce
a mean emitted power per unit solid angle and energy interval along the
line of sight as:

\begin{equation}
<{{dW(r)}\over{dt d\epsilon_1 d\Omega_{SC}}}>=
\int {{D(r,k_3)}\over{\left[\int D(r,k_3) dk_3 \right]}} 
{{dW(r,k_3)}\over{dt d\epsilon_1 d\Omega_{SC}}} dk_3
\end{equation}

where $D(r,k_3)dk_3$ is the volume element at a distance $r$ from the nucleus
with a given $\theta_{SC}$. Under the assumption of an 
ellipsoidal distribution of the relativistic particles 
we find ({\bf Fig.4}):

\begin{equation}
D(r,k_3) = 
\pi -2\,\arcsin\left({{\cos\theta_{Q,r}-k_3 \cos\theta_{ax}}\over
{\sin\theta_{ax}\sqrt{1-k_3^2}}}\right) 
\end{equation}

with

\begin{eqnarray*}
\left| {{ \cos\theta_{Q,r} -k_3 \cos\theta_{ax}}\over
{\sin\theta_{ax} \sqrt{1-k_3^2}}} \right| < 1 \nonumber
\end{eqnarray*}

where $\theta_{ax}$ is the angle between the line of sight and the 
radio-axis and $\theta_{Q,r}$
is the half opening angle at a distance $r$ from the nucleus of the quasar 
emission cone filled by the relativistic particles ($\theta_{Q,r}<\theta_Q$).

The total IC emission has been obtained by integrating
the mean emitted power (B23, B24) 
over all the quasar emission cone 
($\theta_Q = 45^{o}$) under the assumption of an uniform distribution 
of the relativistic electrons.  
The IC emission of a radio lobe along the line of sight is illustrated in 
{\bf Fig. 5}
as function of
$\theta_{ax}$ 
and for three values of $\delta$. This allows a quick estimate of the ratio
of the emissions from the two  lobes as a function of the inclination 
angle.

\begin{figure}
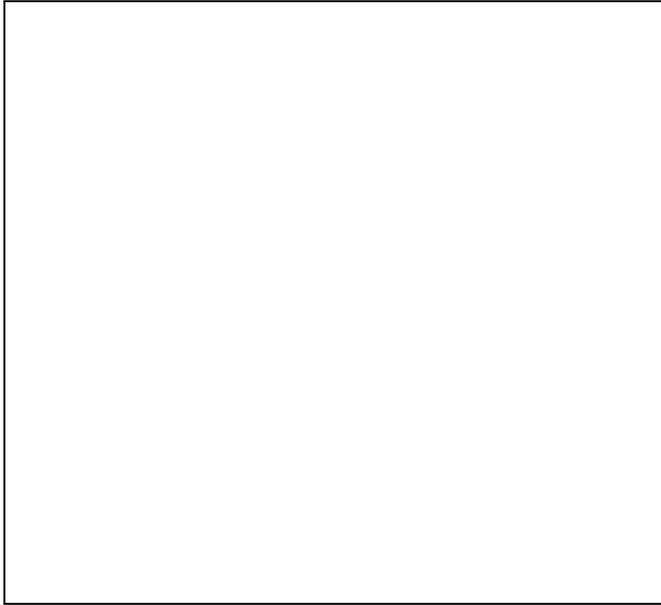

\picplace{8cm}
\caption[]{The IC X-ray luminosity (at 1 keV) of one lobe of a radio galaxy 
as function of the inclination angle $\theta_{ax}$.
The luminosity is normalized to the isotropic IC value 
for the same lobe and is plotted for three representative values of the 
electron spectral index $\delta$:
2.1 (solid line), 2.5 (dotted line) and
3.1 (dashed line).
The computation has been performed assuming a typical ellipsoidal 
configuration 100 $\times$ 50 Kpc 
 and a quasar emission cone half opening angle of 45 degrees.
The ratio between the luminosities of the two lobes can be easily derived
from the plots keeping in mind that the lobes are symmetrically placed 
with respect to $\theta_{ax} =90^o$.}
\end{figure}

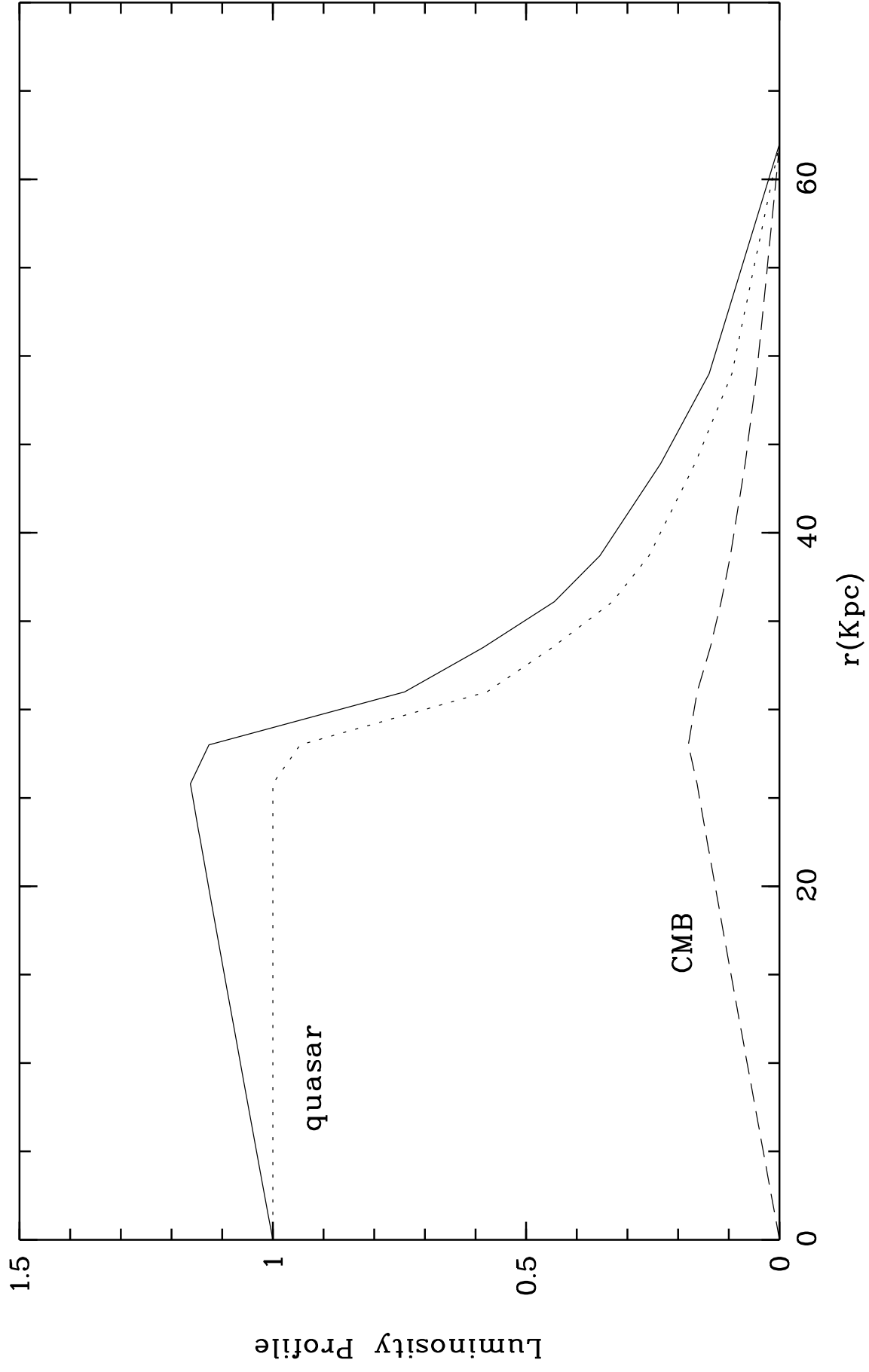
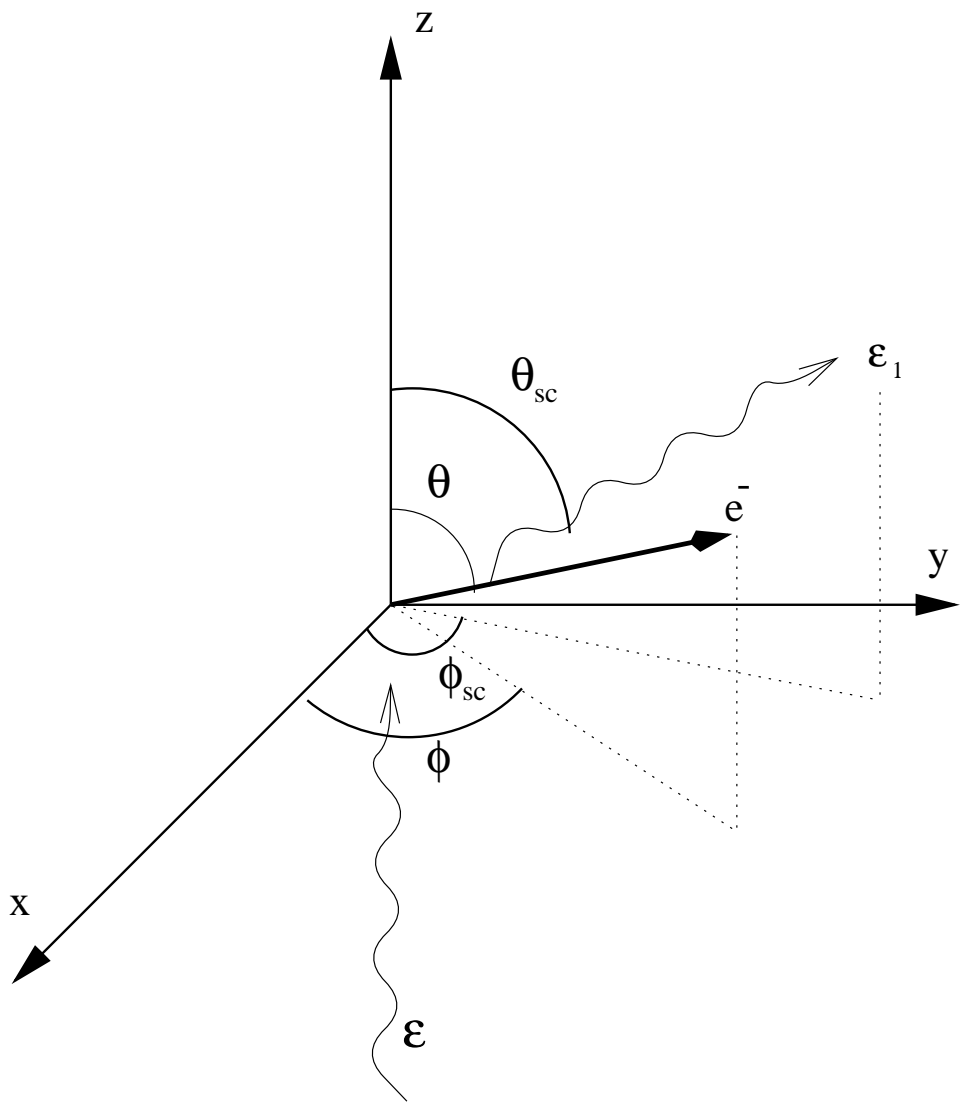
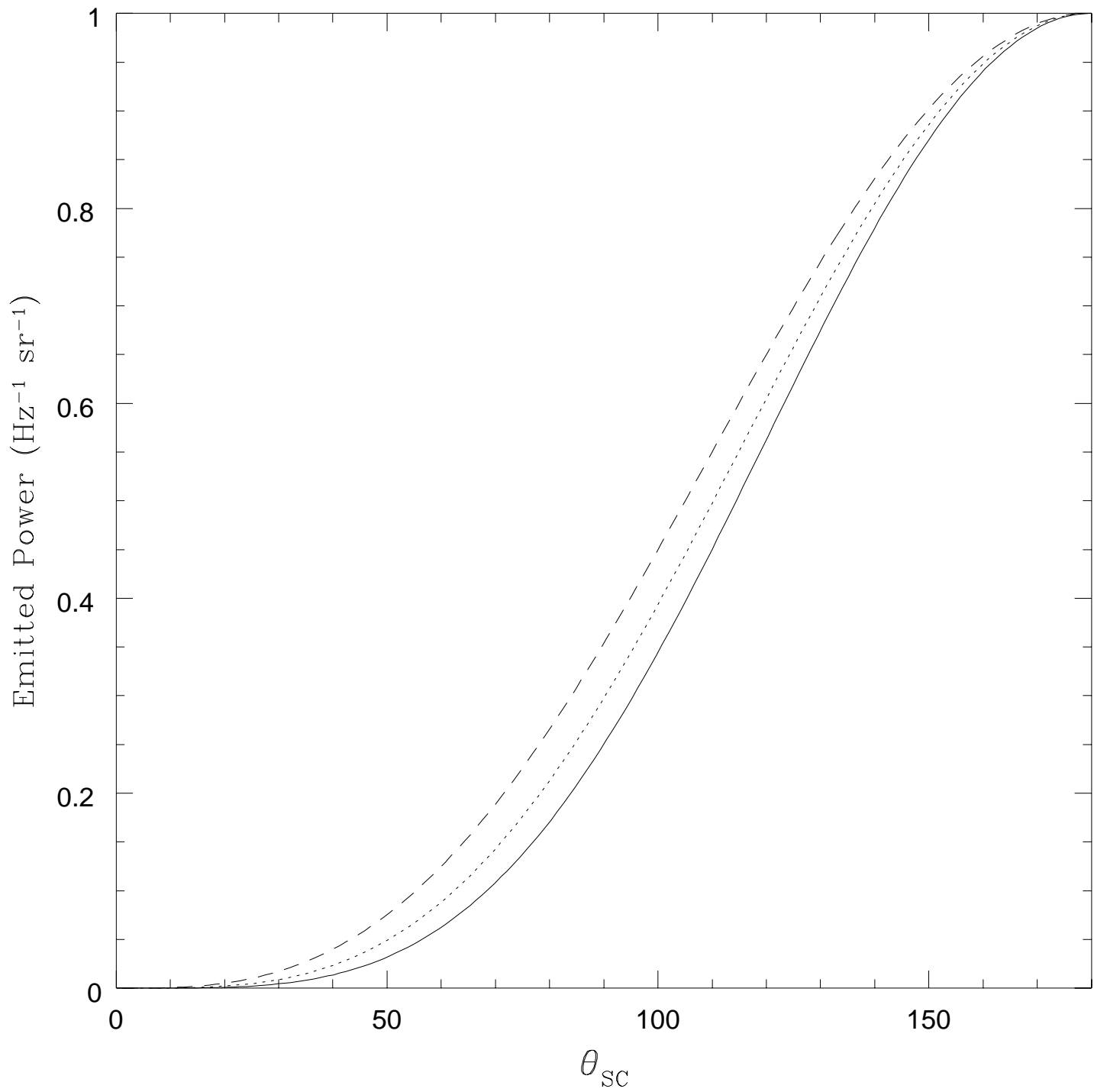
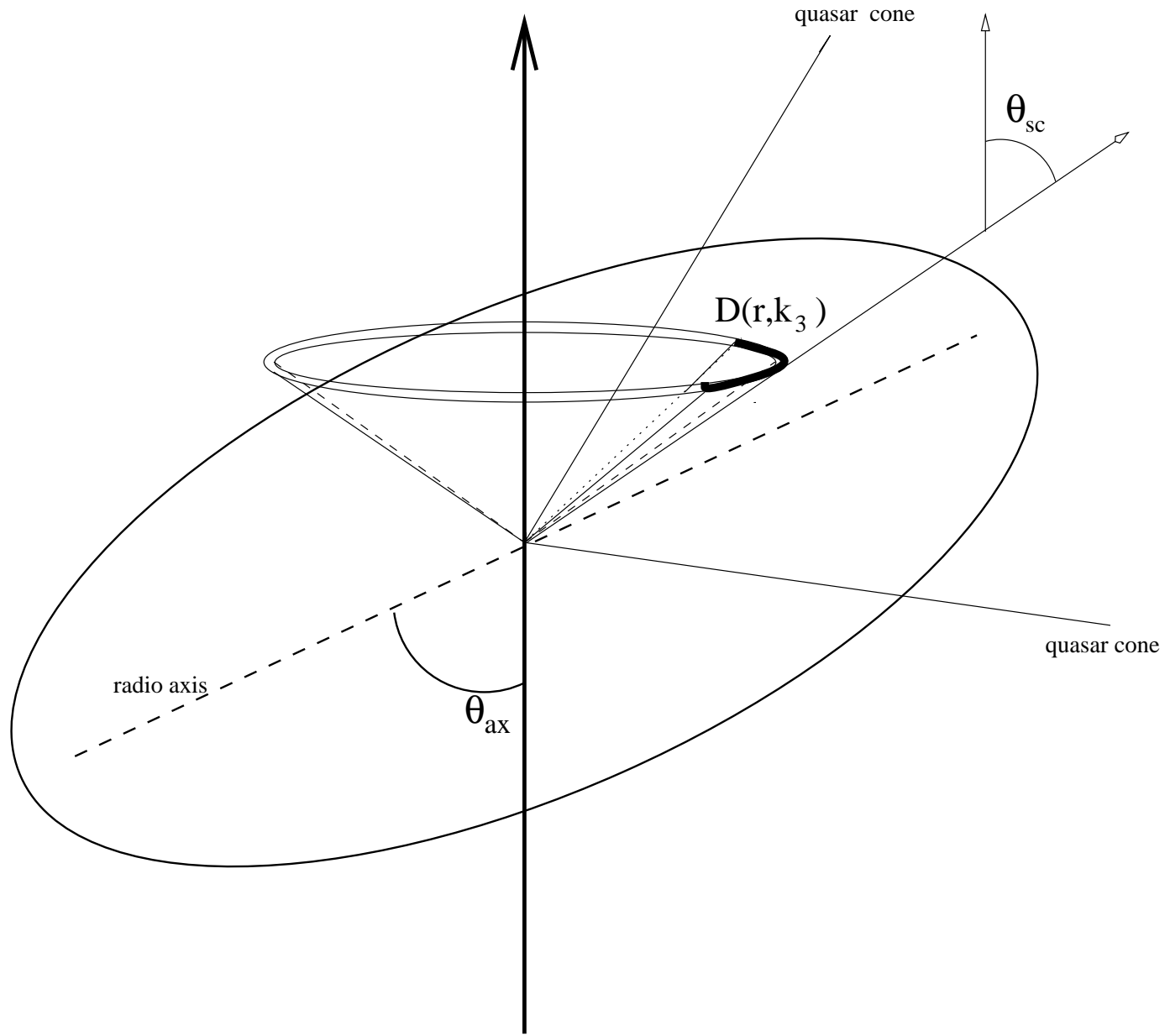
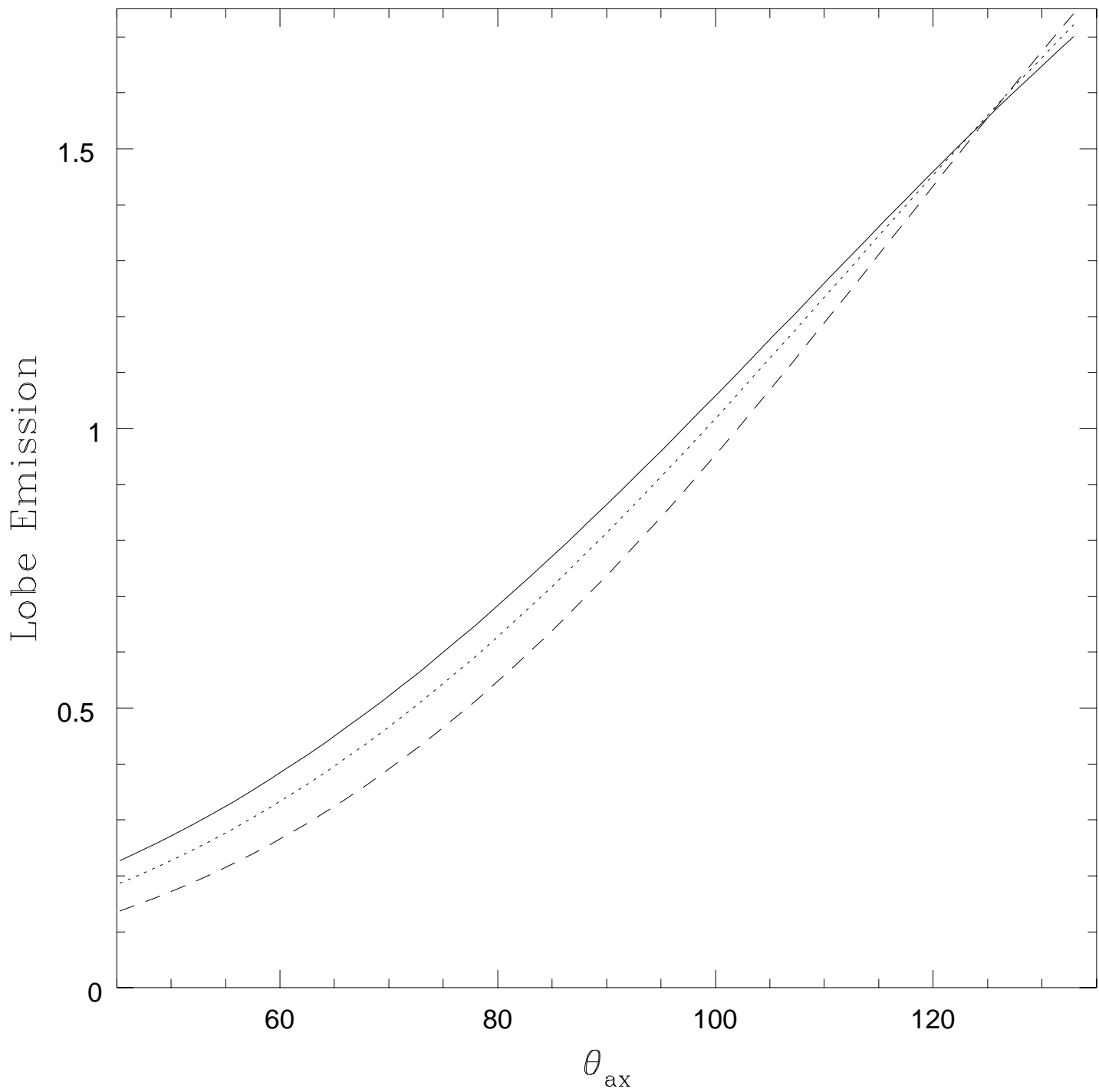
\end{document}